\documentclass{aa}  

\usepackage{graphicx}
\usepackage{txfonts}
\bibpunct{(}{)}{;}{a}{}{,}
\usepackage{hyperref}
\usepackage{booktabs}

\hypersetup{
    colorlinks=true,
    linkcolor=blue,
    urlcolor=blue,
    citecolor=blue,
}
\begin{document}

   \title{X-ray and $\gamma$-ray orbital variability from the $\gamma$-ray binary HESS$~$J1832$-$093}

   \author{G. Martí-Devesa
          \inst{1}
          \and
          O. Reimer\inst{1}
          }

   \institute{Institut f\"{u}r Astro- und Teilchenphysik, Leopold-Franzens-Universit\"{a}t Innsbruck, A-6020 Innsbruck, Austria\\\email{guillem.marti-devesa@uibk.ac.at}}

   \date{Received -; accepted -}

  \abstract
   {Gamma-ray binaries are systems composed of a massive star and a compact object whose interaction leads to particle acceleration up to relativistic energies. In the last fifteen years, a few binaries have been found to emit at high energies, but their number is still low. The TeV source HESS~J1832$-$093 has been proposed as a binary candidate, although its nature is unclear. Neither a GeV counterpart nor a period was detected.}
   { The purpose of this work is to search for a GeV counterpart to understand the origin of the TeV signal detected by HESS. For an unambiguous identification of its binary nature, finding an orbital modulation is crucial.}
   {We analysed data spanning more than 10 years from the \textit{Fermi} Large Area Telescope (\textit{Fermi}-LAT), together with \textit{Swift} archival observations taken between 2015 and 2018, using both the X-Ray Telescope (XRT) and UV/Optical Telescope (UVOT). We searched for periodicities in both X-ray and GeV bands.}
   {We find a periodic modulation of $\sim$ 86 days in the X-ray source candidate counterpart XMMU~J183245$-$0921539, together with indications of $\gamma$-ray modulation with a compatible period in the GeV candidate counterpart 4FGL J1832.9$-$0913. Neither an optical nor a UV counterpart is found at the X-ray source location. The overall spectral energy distribution strongly resembles the known $\gamma$-ray binary HESS~J0632+057.}  
   {Both the spectral energy distribution and the discovery of an orbital period allow the identification of the TeV source HESS~J1832$-$093 as a new member of the $\gamma$-ray binary class.}


   \maketitle
%

\section{Introduction} \label{intro}

The search for binaries emitting $\gamma$-rays resulted in the discovery of a few systems at high energies, the  $\gamma$-ray binaries (see \cite{Dubus13} for a review). These systems are unique because their spectral energy distribution (SED) peaks at $\gamma$-rays, while they show orbitally modulated emission across the electromagnetic spectrum - including their high-energy (HE; E > 100 MeV) and very-high-energy (VHE; E > 100 GeV) components. Almost all the systems known are located in the Galactic plane - \object{PSR~B1259-63} \citep{Aharonian05PSRbin}, \object{LS~5039} \citep{Aharonian06LS5039}, \object{LS~I~+61~303} \citep{Albert06}, \object{HESS~J0632+057} \citep{Aharonian07}, \object{1FGL~J1018.6-5856} \citep{Fermi11binary}, \object{PSR~J2032+4127} \citep{Abeysekara18}, and \object{4FGL~J1405.1-6119} \citep{Corbet19} - with the notable exception of \object{LMC~P3} in the Large Magellanic Cloud (LMC) \citep{Corbet16}.

Gamma-ray binaries are composed of a massive star (O or Be) and a compact object whose nature is in most cases unknown. PSR~B1259$-$63 and PSR~J2032+4127 are the only systems for which a pulsar has been identified, but it is believed that the rest of such binaries may also host a neutron star. It is generally accepted that these systems are direct precursors of high-mass X-ray binaries, before the neutron star enters an accretion state. Their intriguing multi-wavelength nature, together with their periodic behaviour makes them optimal laboratories for studying particle acceleration processes in astrophysical sources - for a review, see \cite{Dubus13} and references therein. Therefore, even if the geometrical conditions for the systems differ significantly, the search for new binaries may contribute to a better understanding of the general features observed and improve the current evolutionary and population models.  One of the main issues within the present status is the number of Galactic $\gamma$-ray binaries whose signal at VHE has not been associated because of their low brightness at HE or other wavelengths - for instance, HESS~J0632+057-like systems \citep{Dubus17}. 

\object{HESS~J1832-093} is an unidentified point-like source located in the Galactic plane, discovered by \cite{HESSCol15} and confirmed in the H.E.S.S. Galactic plane survey \citep{HGPS}. It is partially coincident with the radio shell of the supernova remnant (SNR) \object{G22.7-0.2}. At a distance of $d=4.4\pm0.4$ kpc \citep{Su14}, this SNR might be associated with the $\gamma$-ray signal, since it shows faint non-thermal radio emission and an extension of $26'$ \citep{Shaver70}. Previous systematic studies on SNRs as seen by the \textit{Fermi} Large Area Telescope (\textit{Fermi}-LAT) considered G22.7$-$0.2 as a possible HE emitter in the $1-100$ GeV range \citep{Acero16}. This SNR was initially classified as a candidate for detection, but the strong dependence of the signal significance on the interstellar emission model used for the analysis made its detection uncertain. For this reason, G22.7$-$0.2 was included in the list of non-detected SNRs - see Table 3 in \cite{Acero16}. At VHE, \cite{HESSCol15} argued that the spatial separation between HESS~J1832$-$093 and G22.7$-$0.2, together with the existence of infra-red (IR) and X-ray sources spatially compatible with the $\gamma$-ray signal (\object{2MASS J18324516-0921545} and \object{XMMU~J183245-0921539}, respectively), pointed strongly against an SNR origin. However, the absence of optical and GeV counterparts prevented a clear identification of its nature. Its point-like signal and its significant spatial deviation from the SNR led to ambiguous scenarios for the VHE emission: a pulsar wind nebula (PWN), an active galactic nucleus (AGN) seen through the Galactic plane, a $\gamma$-ray binary or the interaction between protons accelerated in the SNR and a nearby molecular cloud (MC). An interacting MC origin would require slow diffusion, an extended source at TeV and a single hadronic component which should also be seen at HE, while an AGN nature was also disfavoured due to the absence of GeV emission and a soft spectral index at VHE \citep{HESSCol15}. This possibility cannot be completely discarded yet, since it could be an unusual host galaxy. On the other hand, the $\gamma$-ray binary scenario seems to be preferred over the PWN origin because of the variability observed in X-rays \citep{Eger16} and the candidate IR counterpart \citep{Mori17}. Unfortunately, no pulsations from a pulsar have been found, no orbital period has been established, and  no  star  has been found as the optical counterpart. Identifying a GeV source in this multi-wavelength picture could provide crucial information about the origin of the $\gamma$-ray signal. This situation resembles the discovery of HESS~J0632+057, where no periodicity nor HE counterpart was discovered at first, but in later studies \citep{Aharonian07,Bongiorno11,Caliandro13, Li17}.

Motivated by the hints in favour of the binary scenario for HESS~J1832$-$093, we performed a multi-wavelength study of this candidate. In this work, we notice the presence of a HE source close to the binary candidate associated with it, and the SNR shell G22.7$-$0.2 in the 4FGL catalogue \citep{4FGL}. We analyse more than ten years of \textit{Fermi}-LAT data as well as archival X-ray and ultraviolet (UV) data from {\it Swift} (Sections \ref{fermi} and \ref{swift}). Later, we present the spectral results obtained (Section \ref{results-spec}) and the discovery of an orbital period (Section \ref{period}). Finally, we discuss the binary interpretation for XMMU~J183245$-$0921539/4FGL J1832.9$-$0913/HESS~J1832$-$093, and its implications (Section \ref{discussion}), finally summarising our conclusions (Section \ref{summary}).

\section{\textit{Fermi}-LAT observations and data analysis} \label{fermi}


    The LAT is the main $\gamma$-ray detector on board the \textit{Fermi Gamma-ray Space Telescope} \citep{FermiLAT}, covering the energy range between 30 MeV to more than 500 GeV. Its energy-dependent point-spread function (PSF) goes from several degrees at low energies ($\sim 5^{\circ}$ at 100 MeV) to less than $0.1^{\circ}$ above 10 GeV at $68\%$ containment. In this paper, observations from 2008 August 4 to 2018 November 3 were included, using $\sim 10.25$ years of data. The analysis was performed using {\it Fermitools-1.0.1}\footnote{This is the nomenclature for the new \textit{Fermi} Science Tools released through Conda. See \url{https://github.com/fermi-lat/Fermitools-conda/wiki}} on P8R3 reprocessed data \citep{P8R3}. \texttt{SOURCE} event class (evclass=128) and \texttt{FRONT}+\texttt{BACK} event type (evtype=3) were employed, together with the \texttt{P8R3$\_$SOURCE$\_$V2} instrument response functions (IRFs). All photons within a $20^\circ\times20^\circ$ region of interest (ROI) centred on XMMU~J183245$-$0921539 and in the energy range between $100$ MeV and $500$ GeV were selected. Earth limb contamination was handled by selecting events with zenith angles < $90^\circ$. The low energy threshold is motivated by the large uncertainties in the arrival directions of the photons below 100 MeV, leading to confusion between point-like sources and the Galactic diffuse component. See \cite{Principe18} for a different analysis implementation to solve this and other issues at low energies.

Fluxes presented in this work were obtained performing a binned maximum likelihood fit \citep{Mattox96} using \texttt{fermipy 0.17.4}\footnote{A Python package for the $Fermitools$. See \url{https://fermipy.readthedocs.io/en/latest/}} \citep{Fermipy}, with a pixel size of $0.1^\circ$ and $8$ energy bins per decade. The spectral-spatial model employed includes all sources from the 4FGL catalogue \texttt{gll\_psc\_v18} \citep{4FGL} within a $30^\circ\times30^\circ$ region centred on our target. The Galactic and isotropic diffuse components used are 'gll$\_$iem$\_$v07.fits' and 'iso$\_$P8R3$\_$SOURCE$\_$V2$\_$v1.txt', respectively. To evaluate the significance of the detection of each source, the test statistic $TS = -2\;ln\left( L_{max,0}/L_{max,1}\right)$ was used, where $L_{max,0}$ is the log-likelihood value for the null hypothesis (i.e. model without the source) and $L_{max,1}$ the log-likelihood for the complete model. The larger the value of the TS, the less likely $L_{max,0}$ is. A TS of at least 25 ($\sim5\sigma$ evaluated from a $\chi ^2$ distribution with 1 degree of freedom) is required to claim detection of a source \citep{4FGL}.

\begin{figure}
   \centering
   \includegraphics[width=\hsize]{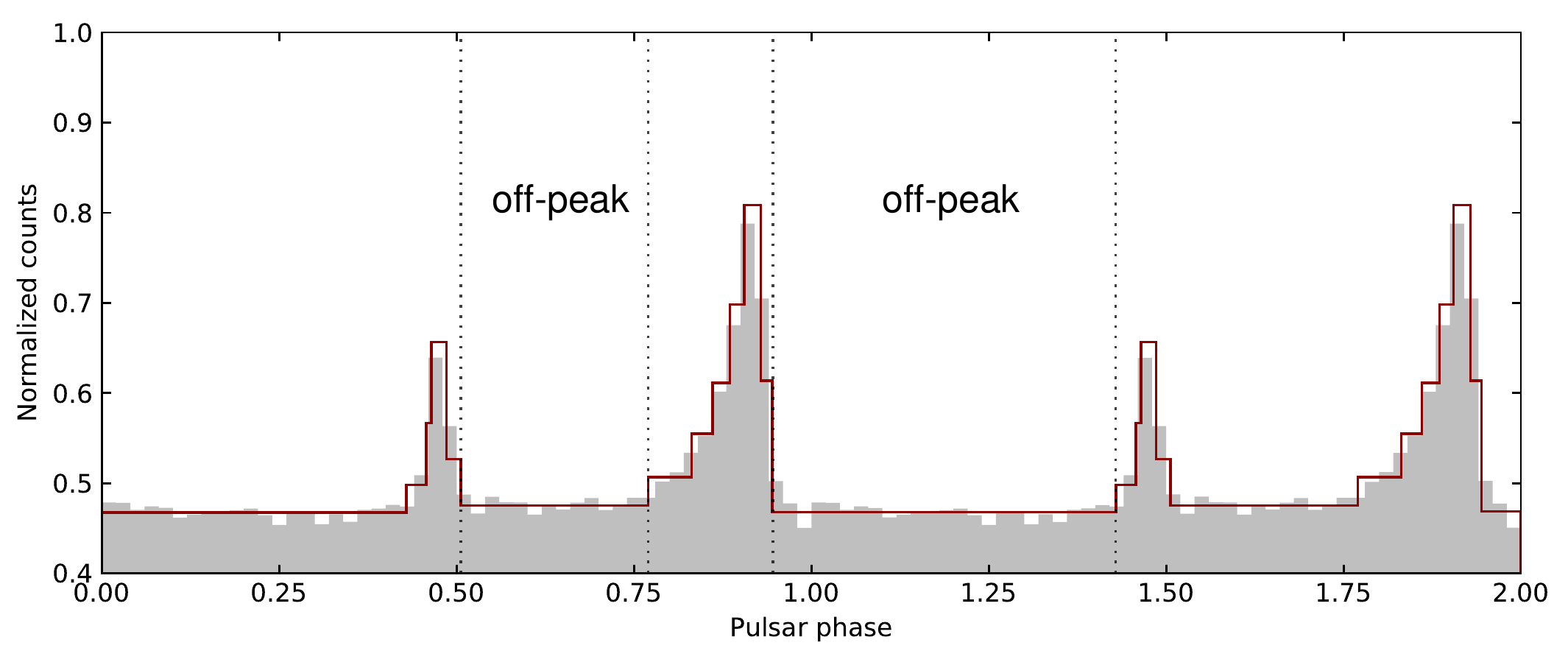}
      \caption{Pulse profile of PSR J1833$-$1034 with an ROI of $1^{\circ}$ above 100 MeV binned with 100 bins per period (grey). Its Bayesian block representation is shown by the red profile. The off-peak phases are delimited by the vertical black lines.
              }
         \label{Fig:pulsar}
 \end{figure} 

The catalogue source \object{4FGL J1832.9-0913} is spatially close to HESS~J1832$-$093, as well as the X-ray and IR counterparts ($\sim 0.14^{\circ}$). We consider it as the possible $\gamma$-ray counterpart of XMMU~J183245$-$0921539. However, 4FGL J1832.9$-$0913 is in a crowded region of the Galactic plane, and contamination from bright pulsars might be significant. We searched for known pulsars close to our target and found \object{PSR~J1832-0836} (at $\sim0.6^{\circ}$) and \object{PSR~J1833-1034} (at $\sim 1^{\circ}$). While the first one is very faint, PSR J1833$-$1034 is bright at low energies \citep{PulsarCat1}. Its emission might be significant and gating it assures no contamination in our ROI. A similar analysis was performed by \cite{Li17} in their study of HESS~J0632+057. Therefore, we split our analysis in two energy bands, gating its emission only below 10 GeV.

In order to attribute a pulsar phase to each event in the ROI, we used \texttt{TEMPO2} \citep{Hobbs06, Edwards06} and the \textit{Fermi} plug-in\footnote{\url{https://fermi.gsfc.nasa.gov/ssc/data/analysis/user/Fermi\_plug\_doc.pdf}} \citep{Ray11}. We adopted an updated ephemeris obtained using the method from \cite{Kerr15}. The pulse profile of PSR J1833$-$1034 for photons within $1^{\circ}$ of the source above 100 MeV is shown in Figure \ref{Fig:pulsar}. Its off-peak phase reported in the second \textit{Fermi}-LAT pulsar catalogue \citep{2PC} does not accurately describe its updated pulse profile. In order to improve it, we decomposed the pulsed light curve using a Bayesian blocks algorithm as detailed in \cite{Scargle13}. This was the method used by \cite{2PC} in the pulsar catalogue. We redefine the off-peak phases as $\Delta \phi = [0.946-0.429]$ and  $\Delta \phi = [0.506-0.770]$. Fluxes were re-scaled by a factor $0.747$ to account for the different exposure time. 

A maximum-likelihood fit was performed on this reduced dataset using the described source model and binning. Firstly, the iterative \texttt{optimize} method from \textit{fermipy} was applied. In a second step, normalization was left free for all sources fewer than $7^{\circ}$ from our target. The smaller PSF at higher energies allowed us to perform an analysis above 10 GeV, with an ROI of $8^\circ\times8^\circ$  and a zenith angle < $105^\circ$. In this case, normalisation for sources within $3^{\circ}$ of 4FGL J1832.9$-$0913 was left free, as well as all parameters for those within $1 ^{\circ}$.

\section{\textit{Swift} observations and data analysis} \label{swift}

The \textit{Neil Gehrels Swift Observatory} \citep{Gehrels04} is one of most versatile missions at X-ray wavelengths. Although its primary goal is to provide detailed information about $\gamma$-ray bursts (GRBs), it is widely used to monitor sources using Target of Oportunity (ToO) observations. This satellite has three instruments on board: the Burst Alert Telescope (BAT) \citep{BAT}, the X-Ray Telescope (XRT) \citep{XRT}, and the UV/Optical Telescope (UVOT) \citep{UVOT}. We used 54 archival on-target observations of XMMU~J183245$-$0921539 with \textit{Swift} (see Appendix \ref{ap-xray}), plus about ten serendipitous observations of the region that might be useful for extracting the flux of our binary candidate. In this paper, only XRT and UVOT data have been used: according to the fluxes previously reported in other works \citep{Eger16,Mori17}, the source is substantially fainter at hard X-rays than the sensitivity limit of the BAT instrument.

\subsection{Swift-XRT analysis} \label{xrt}

The XRT is a grazing-incidence focusing X-ray telescope covering energies between 0.3 and 10 keV, whose ToO programme allows the monitoring of variable X-ray sources. Four observations of XMMU~J183245$-$0921539 with this instrument were included in \cite{Eger16} but have been re-analysed here. For the analysis of the observations, we used the \texttt{HEASOFT v6.26} package with the newest calibration database (\texttt{CALDB}) available. Data were reprocessed using \texttt{XRTPIPELINE}, generating \texttt{CLEANED} level 2 events from grades $0-12$. All PC data from the 54 observations of XMMU~J183245$-$0921539 (i.e. more than 120 ks of exposure) were selected for the spectral extraction. Using \texttt{XSELECT}, a spectrum was obtained from a circular region of $46.6''$ centred on the binary candidate position. For the background, an annular region between $98.8''$ and $188.8''$ was defined. To account for the detector response, a custom effective area file was produced using \texttt{XRTMKARF}, with the \texttt{CALDB}\footnote{Version \href{https://heasarc.gsfc.nasa.gov/docs/heasarc/caldb/data/swift/xrt/index/cif_swift_xrt_20190412.html}{20190412}} file  \texttt{swxpc0to12s6\_20130101v014.rmf}. Additionally, the exposure file generated with \texttt{XRTPIPELINE} was used for the correction. The resulting spectrum was binned with a minimum of 20 counts per energy bin using \texttt{grppha}, and later fitted using \texttt{XSPEC} through its \texttt{pyxspec 2.0.2} interface. A $\chi ^2$ fit was performed assuming an absorbed power-law model with the hydrogen column density $N_H$, the spectral index $\Gamma$, and normalisation as free parameters.

Adding the serendipitous observations, a similar analysis was performed to obtain an unabsorbed light curve between $0.3$ and $10$ keV. All observations were analysed separately, obtaining the spectra using \texttt{XRTPRODUCTS} with the same calibration described above. Due to the low number of counts, we imposed the condition of one count per energy bin with \texttt{grppha}. Finally, the fluxes were obtained from \texttt{XSPEC} performing a C-Stat fit. For this analysis, an absorbed power-law model was assumed again, but fixing  $N_H$ and  $\Gamma$ to the values obtained in the overall spectral analysis described above.

\subsection{Swift-UVOT analysis} \label{uvot}

The UVOT is a diffraction-limited 30 cm Ritchey-Chr\'etien reflector telescope on board \textit{Swift,} with a $17' \times 17'$ field of view with several filters at UV and optical wavelengths. Apart from the XRT observations, the UVOT telescope also observed the location of XMMU~J183245$-$0921539. $V$, $U$, $UVM2$, $UVW1,$ and $UVW2$ filters were used to study the region. In this work, all ToO images with the same filter were summed using \texttt{UVOTIMSUM}. Absolute photometry was applied using both \texttt{UVOTSOURCE} to search for the source and \texttt{UVOTDETECT} to obtain the values for the upper limits. For this purpose, a circular region of $5''$ was selected for the source and an annular region from $8''$ to $13''$ defined for the background to avoid the presence of other sources in the background subtraction. Exposure correction was taken into account.

\begin{figure}
   \centering
   \includegraphics[width=\hsize]{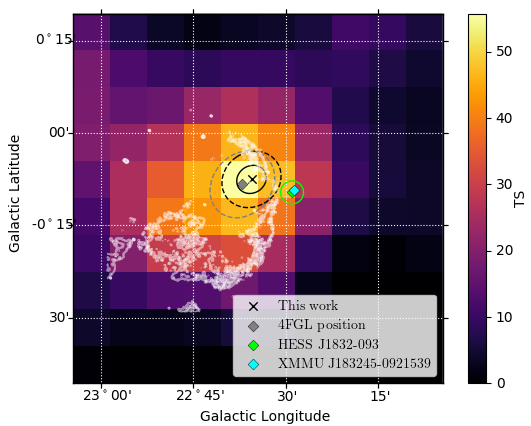}
      \caption{TS map between 100 MeV and 10 GeV of the region around HESS~J1832$-$093 after gating off PSR J1833$-$1034. The map includes the positions of the X-ray, GeV and TeV counterparts, as well as the radio contours of SNR G22.7$-$0.2 \citep{Helfand06}. Confidence contours at 67\% (solid) and 99\% (dashed) levels are added for the 4FGL and the updated source. Additionally, the 95\% confidence contour for the TeV counterpart is included.
              }
         \label{Map}
   \end{figure} 


\begin{table*}
\caption{Summarised spectral properties of HESS~J1832$-$093. Luminosities are computed assuming $d=4.4\pm0.4$ \citep{Su14}. VHE data are from \cite{HESSCol15}}.           
\label{table:allspec}      
\centering                          
\begin{tabular}{c |c c c c c}        
\hline\toprule    
&  Integrated flux &  Luminosity & $\Gamma$ & $\alpha$ & $\beta$\\
\midrule                                    
 VHE (E > 0.1 TeV)& $5.1\pm1.8 \cdot 10^{-12}$ erg cm$^{-2}$ s$^{-1}$&$1.2\cdot 10^{34}$ erg s$^{-1}$ & $2.6\pm0.3$& -&-\\
 HE (0.1--100 GeV) & $1.6\pm0.3 \cdot 10^{-11}$ erg cm$^{-2}$ s$^{-1}$&$3.7\cdot 10^{34}$ erg s$^{-1}$  & -&$2.1\pm0.2$ &$0.435\pm0.007$\\
 X-ray (0.3--10 keV) & $5.9^{+0.3}_{-1.6} \cdot 10^{-13}$ erg cm$^{-2}$ s$^{-1}$ & $1.4\cdot 10^{33}$ erg s$^{-1}$ & $1.46 \pm 0.33$&- &-\\

\end{tabular}
\end{table*}

\section{Spectral results} \label{results-spec}

In this section, we present a study of the potential binary spectrum in different energy bands, from the UV to the HE regime.

\subsection{GeV energies} \label{results-gev}

4FGL J1832.9$-$0913 is a $\gamma$-ray source included in the 4FGL catalogue \citep{4FGL}, close to the position of HESS~J1832$-$093, and spatially compatible with the SNR G22.7$-$0.2. It is detected in the catalogue at $7.6\sigma$, with an integrated energy flux between 100 MeV and 100 GeV of $1.88\pm0.72\cdot10^{-11}$ erg cm$^{-2}$ s$^{-1}$ (here and hereafter, uncertainties are defined at $1\sigma$ confidence level). Its spectrum is modelled with a log-parabola,

\begin{equation}
                \frac{dN}{dE}=N_0\left( \frac{E}{E_b} \right)^{-\left( \alpha+\beta\log \left( E/E_b\right) \right),}
        \end{equation}

where $E_b$ is a scale parameter fixed to 1.8 GeV. While it is associated in the catalogue with the SNR and HESS~J1832$-$093, 4FGL J1832.9$-$0913 is located  only $0.14^{\circ}$ from the position of the X-ray counterpart (slightly beyond the $95\%$ confidence limit - see Figure \ref{Map}). This source is qualified with Flag 2, meaning its position moves beyond the $95\%$ ellipse when changing the diffuse model \citep{4FGL}.

\begin{figure}
   \centering
   \includegraphics[width=\hsize]{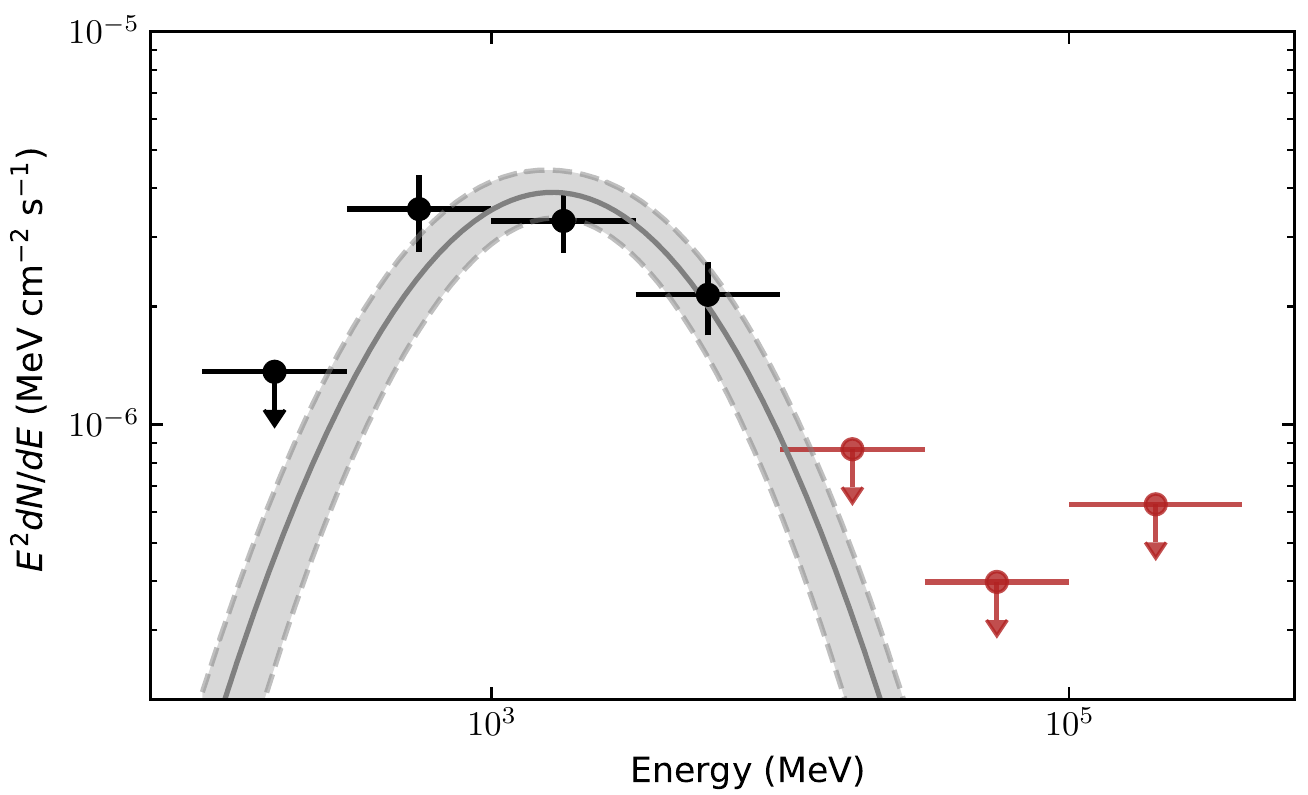}
      \caption{SED for 4FGL J1832.9$-$0913 (black) and best fit (grey) from the analysis below 10 GeV after gating off pulsed emission from PSR J1833$-$1034. In red, upper limits from the analysis above 10~GeV.
              }
         \label{fermi-sed}
   
\end{figure} 

Following the analysis described in Section \ref{fermi} (i.e. two energy bands, PSR J1833$-$1034 gated), 4FGL J1832.9$-$0913 is detected with $TS = 85.72$ ($9.26\sigma$) below 10 GeV. We evaluated its spectrum performing an extra fit with all its spectral parameters free, as well as the normalisation of sources within $3^\circ$ of our target. An integrated energy flux of $1.60\pm0.25\cdot10^{-11}$ erg cm$^{-2}$ s$^{-1}$ (Table \ref{table:allspec}) is obtained between 100 MeV and 10 GeV, with $\alpha=2.1\pm0.2$ and $\beta=0.435\pm0.007$. Its SED can be seen in Figure \ref{fermi-sed}, and it is compatible with the results from the 4FGL catalogue. Additionally, we refined its position employing the \texttt{localize} function from \textit{fermipy}. The updated location of the $\gamma$-ray signal is ($l,b$) = ($22.593^\circ\pm0.026^\circ,-0.125^\circ\pm0.025^\circ$), reducing the distance between the X-ray and GeV sources to $0.12^{\circ}$.

The analysis above 10 GeV using the whole dataset and a zenith angle $>105^{\circ}$ does not yield significant emission from the source ($TS=3.1$), leading to an upper limit of $6.13\cdot10^{-11}$ ph cm$^{-2}$ s$^{-1}$ at $95\%$ confidence level. This result confirms the non-detection of any $\gamma$-ray source with \textit{Fermi}-LAT in \cite{HESSCol15}, when an analysis above 10 GeV was performed with only four years of data.

\subsection{UV and X-ray wavelengths} \label{results-xrt}

\begin{figure}
   \centering
   \includegraphics[trim={0.5cm 1.5cm 3.05cm 2.5cm},clip,width=\hsize]{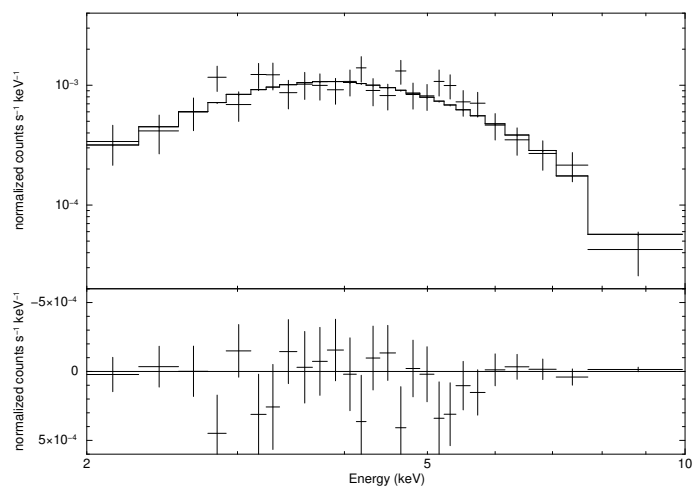}
      \caption{\textit{Top}: X-ray spectrum for XMMU~J183245--0921539 using all observations centred on the binary, fitted to an absorbed power law and zoomed between 2 and 10 keV. \textit{Bottom}: Residuals of the fit.
              }
         \label{XRT_spectrum}
   \end{figure} 

XMMU~J183245$-$0921539 spectrum at keV energies is better described by a power law than using a single thermal component \citep{Eger16}. The best fit with \texttt{XSPEC} to the archived data between 0.3 and 10 keV provides an unabsorbed integrated energy flux of $5.9^{+0.3}_{-1.6} \cdot 10^{-13}$ erg cm$^{-2}$ s$^{-1}$, a column density $N_H= (7.3 \pm 1.4) \cdot 10^{22}$ cm$^{-2}$ and a spectral index $\Gamma=1.46 \pm 0.33$ (Figure \ref{XRT_spectrum}). Both $N_H$ and $\Gamma$ are fully compatible with previous results \citep{Mori17}, but the flux is lower. From a binary system, we expect variability along an orbit, thus this value would correspond to an average flux if the orbit is completely covered. 

The analysis from UVOT data provides for the first time upper limits of the source at UV wavelengths (see Appendix \ref{ap-xray}). The non-detection at optical wavelengths (filter $V$) down to $2.03 \cdot 10^{-17}$ erg cm$^{-2}$ s$^{-1}$ $\r{A}^{-1}$ at 95\% confidence level constrains the flux of the companion star to lower levels than the previous upper limits from surveys (see Section \ref{optic}).

\section{Search for periodicity} \label{period}

In order to confirm the binary nature of XMMU~J183245$-$0921539 and unambiguously associate it to 4FGL J1832.9$-$0913, we need to find  an orbital period. Therefore, we searched for a periodic modulation of their flux.

\subsection{X-ray orbital modulation}

\begin{figure}
   \centering
    \center
    \includegraphics[width=\linewidth]{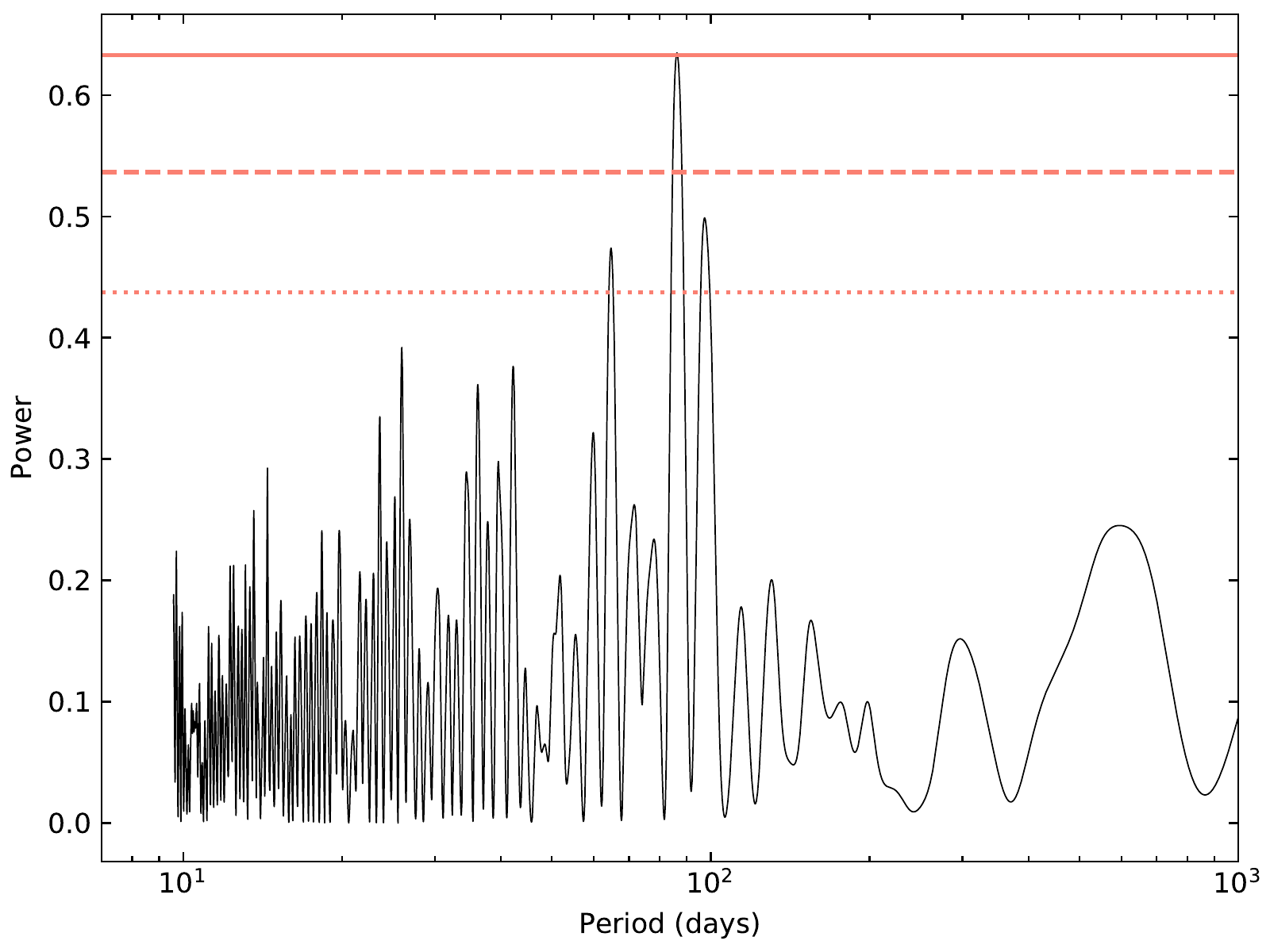}
    \caption{Lomb-Scargle periodogram for the unbinned light curve in X-rays. A peak is found at $P = 86.28 \pm 3.77$ days. FAP levels are represented in red at $3\sigma$ (dotted), $4\sigma$ (dashed), and $5\sigma$ (solid).
              }
         \label{xray-periodograms}
   \end{figure}

\begin{figure}
   
    \center
    \includegraphics[width=\linewidth]{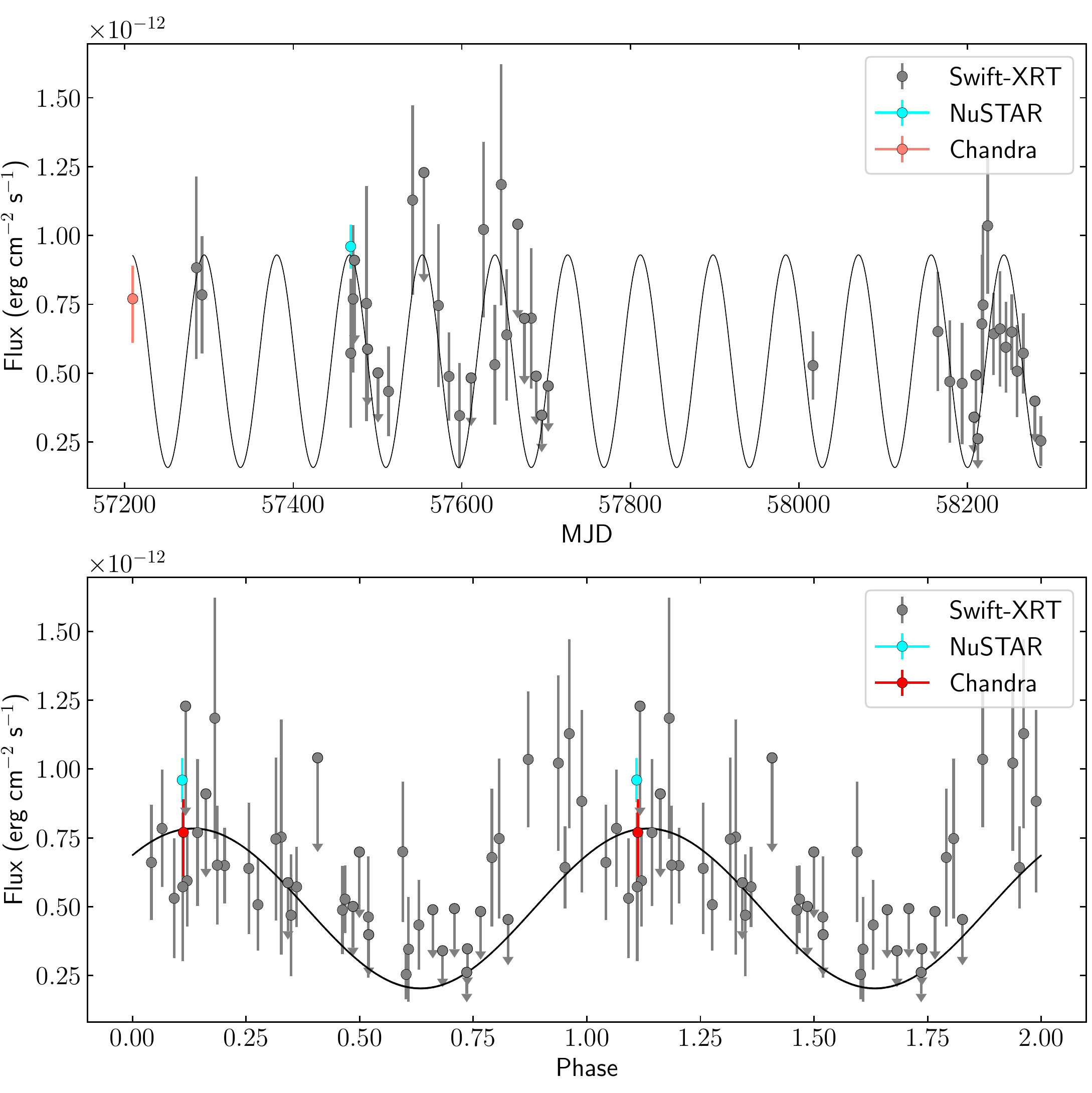}
   
    \caption{\textit{Top:} X-ray light curve for XMMU~J183245$-$0921539 in time space, including all observations performed after 57200 MJD. A sinusoidal function with fixed period to 86.28 days is fitted to the data. \textit{Bottom:} Phase-folded light curve of the same observations with that period. XMM re-analysis from \cite{Mori17} has been added for completeness. Data are duplicated in two orbits for visualization purposes.
              }
   \label{fig:lc-xrays}
\end{figure}

X-ray orbital modulation has been observed for all the known $\gamma$-ray binaries. \cite{Eger16} suggested, for the first time, X-ray variability from XMMU~J183245$-$0921539; however, this assessment is insufficient to establish the binary nature without periodicity. Using the observations analysed in this work and the fluxes reported by \cite{Mori17}, we searched for a periodic variability through a Lomb-Scargle periodogram \citep{Lomb76,Scargle82}. To reduce the noise, only observations with more than 1ks of exposure were selected, and observations earlier than 57200 MJD were discarded due to their poor sampling of the time space. A peak is found at $86.28 \pm 3.77$ days, with a false alarm probability (FAP) smaller than 5$\sigma$ (Figure \ref{xray-periodograms}). This FAP is computed using the analytical approximation from \cite{Baluev08} implemented in \textit{astropy} \citep{astropy:2018}. The peak is still detected with $p < 0.01$ using the more conservative bootstrap method \citep{VanderPlas18}. We arbitrarily define $\phi=0$ at $T_0$ = 54524.9979255 (the oldest observation performed with \textit{Swift}-XRT of the source). Phase-folding the light curve with this period shows orbital variability (Figure \ref{fig:lc-xrays}). We can see how the \textit{NuSTAR} and \textit{Chandra} observations \citep{Mori17} correspond to the peaks of the light curve. Similar results are found if the \textit{Swift} observations are binned assuring at least 1.5 and 2.5 ks per bin, with peaks at $85.94\pm 3.94$ and $85.40\pm 3.93$ days, respectively. Similar background subtractions to the one described in Section \ref{xrt} have been tested and provide compatible results. Failure modes or other defects \citep{VanderPlas18} cannot describe the signal found. Additionally, an epoch-folding method is employed \citep{Leahy87}. This was successfully used to establish the period of the binary 1FGL J1018.6$-$5856 with \textit{Swift} observations \citep{An13}. This methodology allows us to use all observations regardless of individual exposures. The broad peak found is compatible with the previous result, but it is not an improvement compared with the Lomb-Scargle algorithm.

Two other non-negligible peaks are found in Figure \ref{xray-periodograms}, at $\sim 96$ and $\sim 65$ days, respectively, and both of them have a non-physical origin. The first one can be explained as a resonance of the window function, meaning it is an artefact produced by the sample of times $t_i$ of the observations. The second one is an effect of the deviation from a sinusoid of the light curve. If a second mode is added to the regular Lomb-Scargle periodogram, a peak is found at $85.37$ days and the light curve is well represented by the fit. In this case, no peak is found at $\sim 65$ days, confirming its origin as an artefact of the residual. Gamma-ray binaries show a non-symmetric peak in their X-ray light curve peaks \citep{Bongiorno11,Dubus15,Barkov18,Corbet19}; thus we do not expect an exact sinusoidal modulation. 

\subsection{Search for GeV orbital modulation}

If 4FGL J1832.9$-$0913 is associated with XMMU~J183245$-$0921539, a similar periodicity should be found in the $\gamma$-ray data. Period searches using Fourier series in {\it Fermi}-LAT data are described by \cite{Corbet10}, and blind searches using this method on $\gamma$-ray sources led to the discovery of the $\gamma$-ray binaries LMC P3 and 4FGL J1405.1$-$6119 \citep{Corbet16,Corbet19}. Using all events within $2^{\circ}$ of the binary candidate, we assigned a probability to each of them using the model obtained from the analysis described in Section \ref{fermi}. Time bins of 1500 ks were used and weighted with the exposure. No signal is found in the power spectrum obtained. Similar results are obtained if a larger binning is used (e.g. three days). 

\begin{figure}
   \centering
   \includegraphics[width=\hsize]{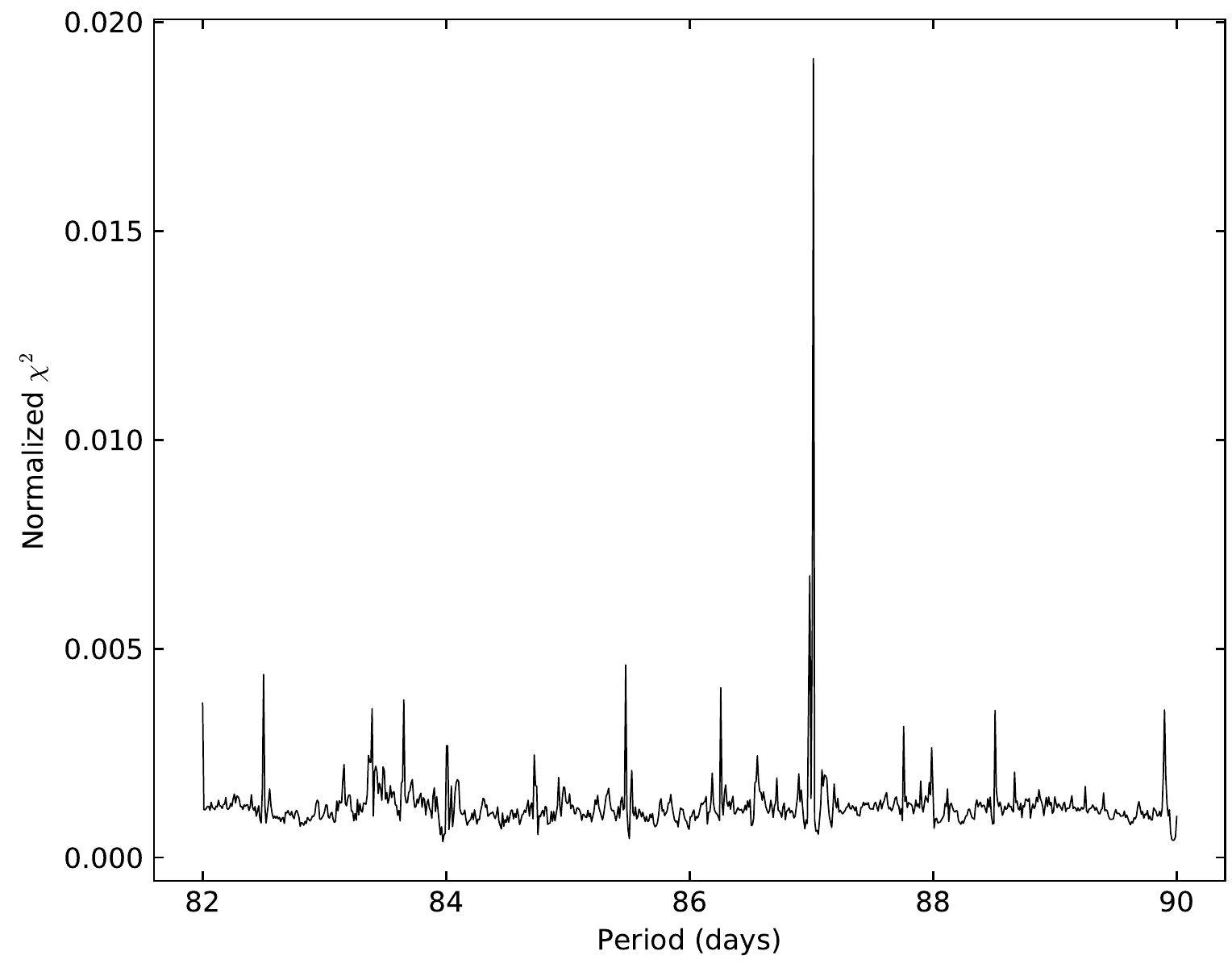}
      \caption{$\chi^2$ vs period for the {\it Fermi}-LAT data. A peak is found at 87.016 days.
              }
         \label{chi2-fermi}
   \end{figure}

Additionally, we explored the possibility of adapting the epoch-folding method mentioned before to the $\gamma$-ray data. \cite{Leahy87} proposed an epoch-folding method to determine the period $P$ and the amplitude $A$ of signals with a sinusoidal nature. All the data are phase-folded for a grid of periods $P'$ and binned in the phase space in $n$ bins. Finally, all phase-folded light curves are compared with the average flux using a $\chi^2$ test to search for variability, and those values are plotted in a $\chi^2$ vs $P'$ plot. This plot should show a peak for the true value of $P$ and white noise for other $P'$. This method contemplates a $\chi^2$ test for count rates, but we can use the summed probabilities per bin weighted with the exposure, using $n=10$. A peak is found at 87.106 days (see Figure \ref{chi2-fermi}).  Unfortunately, the \textit{S} statistic defined in \cite{Leahy87} does not converge, and varying the number of bins used also enhances the smaller peaks found nearby, reaching similar levels. This makes the result uncertain and might indicate the existence of variability, but it cannot confirm the true period. 

\begin{figure}
   
    \center
    \includegraphics[width=\linewidth]{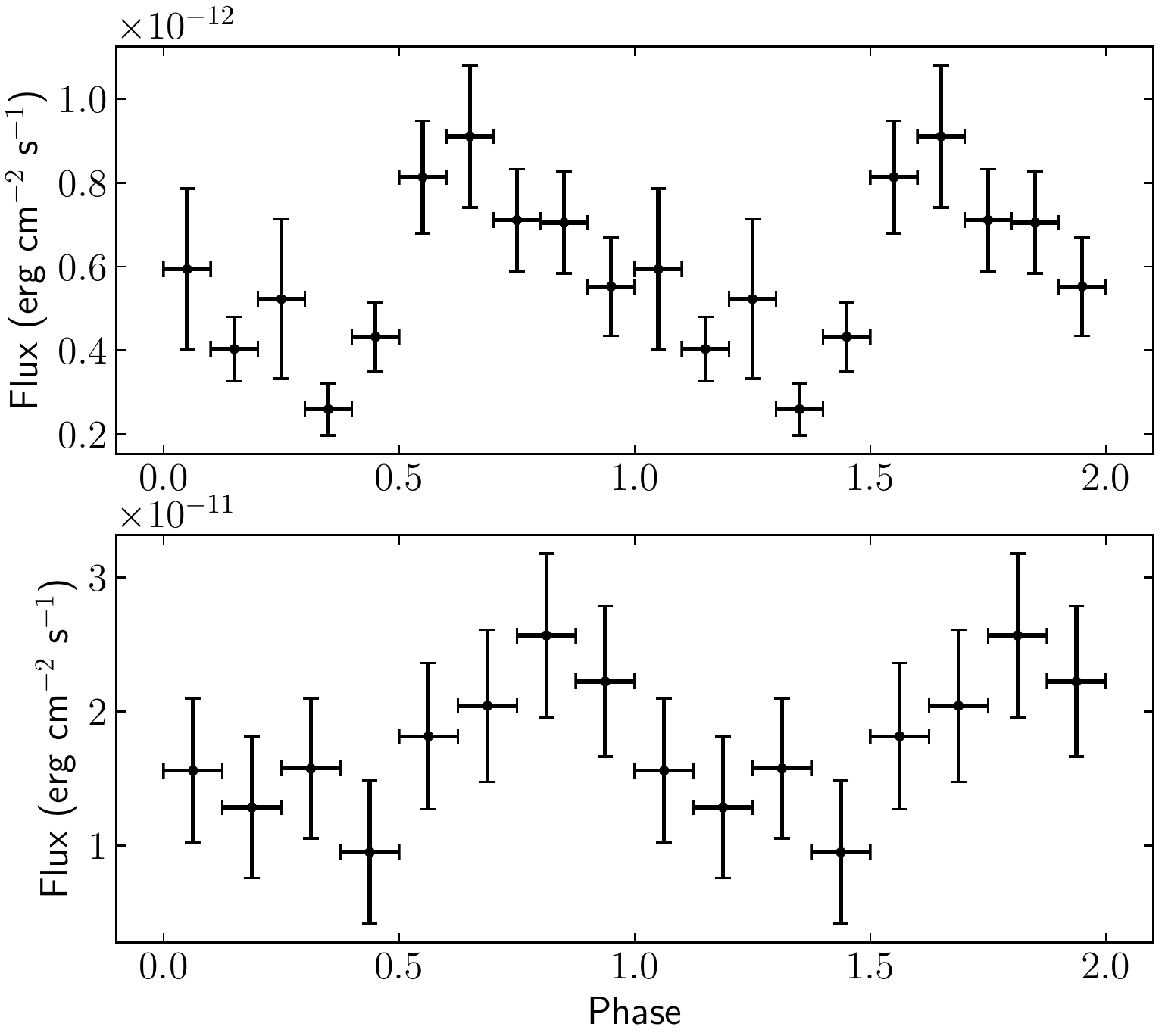}
   
    \caption{Light curves folded with $P=87.016$ days. \textit{Top:} X-rays from \textit{Swift}-XRT. \textit{Bottom:} \textit{Fermi}-LAT data. All fluxes are significantly detected above $2\sigma$ per bin. Data are duplicated in two orbits for visualisation purposes.
              }
   \label{fig:lc_gev}
\end{figure}

These methods for period searches make use of the aperture photometry analysis using all events. A different approach would be to explore the variability of the system performing a maximum likelihood analysis in a phase-binned analysis. We phase-folded the $\gamma$-ray data with the 87.106-day period. When an analysis is performed as described in Section \ref{fermi}, variability can be appreciated (Figure \ref{fig:lc_gev}). However, due to the large flux uncertainties, a $\chi^2$ test cannot discard the hypothesis of non-variability ($\chi^2_{d.o.f.}\approx1$). Similar results are found when phase-folding the data with the periods obtained using the binned light curves in X-rays. Searches for variability by phase-folding the results with periods much larger or smaller than $86 \pm3 $ days do not indicate any variability. 

\section{Discussion} \label{discussion}

\begin{figure*}[h]
   \centering
   \includegraphics[width=\hsize]{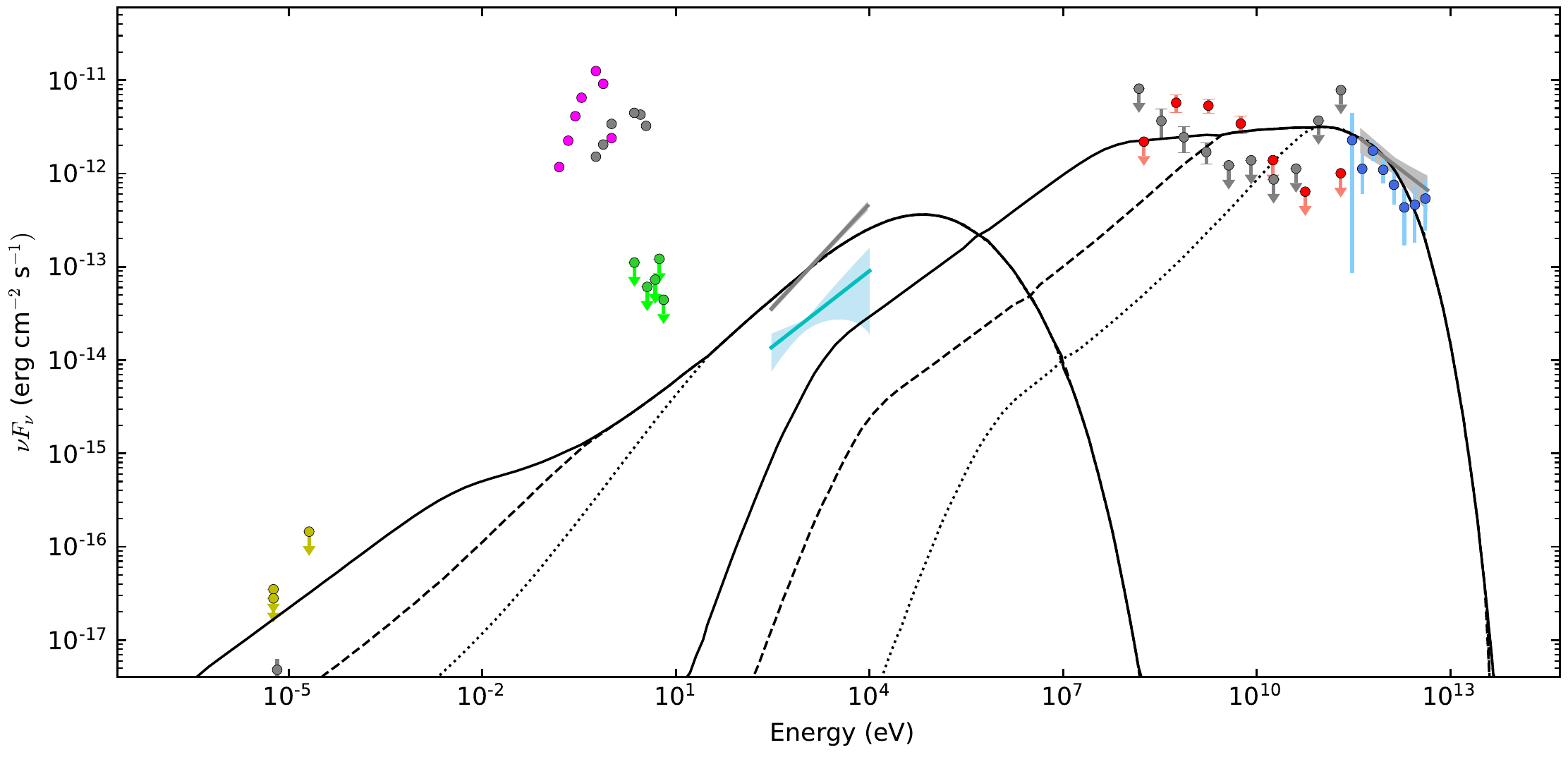}
      \caption{SED of HESS~J1832$-$093: radio upper limits (yellow) \citep{Condon98,Helfand06}, IR (magenta) \citep{2MASS}, UV (green), X-rays (cian), HE (red), and VHE (blue) $\gamma$-rays \citep{HESSCol15}. In grey, SED from HESS~J0632+057 is shown for comparison, including radio \citep{Moldon11}, IR re-scaled a factor $10^{-3}$ \citep{2MASS}, X-ray \citep{Hinton09}, and $\gamma$-rays \citep{Li17,Aharonian07}. In black, synchrotron and IC emission models for HESS~J0632+057 with $E_{min}$ of 1 GeV (solid), 10 GeV (dashed) and 100 GeV (dotted) \citep{Hinton09}.
              }
         \label{SED}
\end{figure*}

Given the spatial coincidences of the sources at different wavelengths (Figure \ref{Map}) and the flux variability at GeV and keV energies (Figure \ref{fig:lc_gev}), the association between XMMU~J183245$-$0921539, 4FGL J1832.9$-$0913 and HESS~J1832$-$093 is reinforced. Its composite SED using the results obtained in previous sections together with the VHE data \citep{HESSCol15,HGPS} peaks in $\gamma$-rays (Figure  \ref{SED}). Radio upper limits are obtained from MAGPIS \citep{Helfand06} and NVSS \citep{Condon98}. Interestingly, the TeV and GeV components do not arise from a single PL, since the upper limits at intermediate $\gamma$-rays prevent such connection. This behaviour is typically observed in $\gamma$-ray binaries, where a cut-off between GeV and TeV components is present. This phenomenology showing two separate components has been understood as synchrotron and inverse Compton (IC) emission produced by two different populations of electrons accelerated in the shocked pulsar wind and the Coriolis turnover,\\  respectively \citep{Zabalza13}, being a consequence of the orbital motion. Actually, the SED observed in Figure \ref{SED} strongly resembles the spectrum from HESS~J0632+057, with an excess in the GeV component that might arise from particularities of HESS~J1832--093, like the geometry of the system, or a contribution from the nearby SNR G22.7$-$0.2.

Regarding the HESS~J1832$-$093 orbital variability, the hints of periodicity in $\gamma$-rays allow a comparison between both light curves. In Figure \ref{fig:lc_gev}, we see how the peak shifts to later phases from X-ray to GeV. This is a common phenomenon in $\gamma$-ray binaries \citep{Chang16}. Therefore, the relation between GeV and TeV sources for HESS J1832$-$093 is strengthened. On the other hand, for all $\gamma$-ray binaries known, the X-ray and TeV light curves are correlated \citep{Dubus13,Aliu14}. Thus, we expect the TeV component to peak at the same phase as in X-rays. However, it should be noted that these relations between $\gamma$-ray and X-ray light curves might change over several orbits for some binary systems \citep{Hadasch12}.

\subsection{A new GeV-faint $\gamma$-ray binary} \label{population}

In light of the new results presented in this work, we can finally identify HESS~J1832$-$093 as a new member of the $\gamma$-ray binary class. \cite{Corbet16} suggested that the discovery of LMC P3 in the LMC could be interpreted as an indication that almost all the observable population of $\gamma$-ray binaries in our Galaxy had been already discovered. However, the confirmation of the binary nature of HESS~J1832$-$093, and the recent discovery of 4FGL J1405.1$-$6119 \citep{Corbet19}, do not support this interpretation.

Early works predicted an overall population of $\sim 30$ $\gamma$-ray binaries in the Milky Way \citep{Meurs89}. However, in a more recent study performed by \cite{Dubus17}, the population of Galactic $\gamma$-ray binaries was estimated to be $101^{+89}_{-52}$. The main factor of uncertainty in this model is the TeV unassociated sources whose GeV component might be faint (i.e. systems similar to HESS~J0632+057 and HESS~J1832-093). According to \cite{Dubus17}, such systems have extremely low probabilities of being detected in \textit{Fermi}-LAT or H.E.S.S.-like surveys ($\sim 0.8\%$), but the detection of HESS~J0632+057 as a faint GeV source \citep{Li17} placed an upper limit to the number of similar systems as 231. 

The resemblance between HESS~J1832$-$093 and HESS~J0632+057 might be an indication of a population of $\gamma$-ray binaries which have not been identified due to observational biases. While other steady binaries have $L_{GeV}\approx10^{35}$ erg s$^{-1}$,  HESS~J1832$-$093 is one order of magnitude fainter at HE but similarly bright at VHE - see Table 2 in \cite{Dubus13}. The detection threshold in systematic searches for periodicity performed by \cite{Corbet19} is consistent with the non-identification of HESS~J1832$-$093 due to its lack of detection above 10 GeV. Given the low probability of serendipitous detections of new binaries, new observational approaches are required. Multi-wavelength synergies between observatories at all frequencies are necessary, especially between X-rays and VHE. \cite{Dubus17} estimated that the CTA \citep{CTA19} will be able to detect a few of these systems within the first two years of observations, but as is seen in the case of HESS~J1832$-$093, X-ray surveys such as {\it eROSITA} \citep{eRosita12} could be used to properly identify the systems once they are detected.

\subsection{On the optical counterpart issue} \label{optic}

Apart from period searches, resolved spectra from a faint source would unambiguously distinguish between a binary containing a massive star or an AGN. Unfortunately, no optical counterpart is found in the sky surveys performed at visible wavelengths, even in {\it Gaia} DR2\footnote{\url{https://gea.esac.esa.int/archive/documentation/GDR2/} \citep{Gaia16}}. After finding an orbital modulation at other wavelengths, the absence of an optical counterpart has to be properly understood. 

Since G22.7$-$0.2 is in a complex region with MCs, H$_{II}$ regions, and the \object{GLIMPSE9} stellar cluster \citep{Su14}, we consider the possibility of having a binary object related with those systems, thus at a distance of $d=4.4\pm0.4$ kpc. Assuming this distance and using previous observations from the literature of the IR candidate counterpart 2MASS J18324516$-$0921545, \cite{Mori17} derived a spectral type between B8V and B1.5V, fully compatible with the stars found in other $\gamma$-ray binaries \citep{Dubus13}. They obtained an optical absorption of $A_V=7.7$ using the hydrogen column density from radio surveys $N_H= 1.7 \cdot 10^{22}$ cm$^{-2}$\citep{Gorenstein75, Willingale13, Mori17}. Using this extinction and the typical luminosities of spectral types B0-B8 \citep{Binney-Book}, we obtain a range of $V$ magnitudes between 16.7 and 20.7. We note that this is only a lower limit, since the hydrogen column density obtained from X-rays is significantly higher ($N_H= 7.3 \cdot 10^{22}$ cm$^{-2}$). The result is consistent with its non-detection, taking into account that {\it Gaia} DR2 completeness is affected for sources fainter than $G=17$ magnitudes due to systematics, especially in crowded regions in the Galactic plane.

However, our upper limit in the $V$ filter of 20.5 magnitudes is more restrictive, and the source should have been observed. Therefore local dust absorption is required if the assumption of $d=4.4\pm0.4$ kpc is correct. As \cite{Mori17} suggested, high absorption would also allow an O primary star, since we obtain an upper limit for the optical absorption of $A_V=33.2$ using the X-ray measurement. New data releases from {\it Gaia} or dedicated photometric and spectroscopic observations are necessary to identify the stellar type of the companion and obtain a proper distance to the system.

\section{Summary} \label{summary}

We used \textit{Fermi}-LAT data and \textit{Swift} archival observations to understand the origin of the TeV source HESS~J1832--093. The spectral and temporal analysis performed in the present work led to the following results:

   \begin{enumerate}
      \item 4FGL J1832.9$-$0913 is a $\gamma$-ray source spatially compatible with the binary candidate HESS~J1832$-$093. We update its position, reducing the distance to $0.12^{\circ}$, which is smaller than our PSF at 1 GeV ($0.8^{\circ}$ at $68\%$ containment).
      \item This source is only detected below 10 GeV with \textit{Fermi}-LAT. 
      \item A period of $\sim 86$ days is obtained from the X-ray {\it Swift}-XRT data, confirming the existence of a binary system. Indications of a similar periodicity are found in the  {\it Fermi}-LAT $\gamma$-ray data.
      \item The SED shows a bimodal component at high energies, a feature characteristic of $\gamma$-ray binaries. In particular, the SED from HESS~J1832$-$093 strongly resembles that of the binary HESS~J0632+057.
      \item The population of $\gamma$-ray binaries might be larger than expected due to the existence of further faint-GeV binaries like HESS~J1832$-$093 and HESS~J0632+057. 
   \end{enumerate}

HESS~J1832--093 is therefore identified as a new $\gamma$-ray binary. Subsequent multi-wavelength observations will be able to establish the geometrical, thermal and high-energy properties of the system. \footnote{During the publication process of this work, \cite{Tam20} reported a multi-wavelength study of HESS~J1832--093.}

\begin{acknowledgements}
The {\it Fermi}-LAT Collaboration acknowledges generous ongoing support from a number of agencies and institutes that have supported both the development and the operation of the LAT as well as scientific data analysis. These include the National Aeronautics and Space Administration and the Department of Energy in the United States, the Commissariat \`{a} l'Energie Atomique and the Centre National de la Recherche Scientifique / Institut National de Physique Nucl\'{e}aire et de Physique des Particules in France, the Agenzia Spaziale Italiana and the Istituto Nazionale di Fisica Nucleare in Italy, the Ministry of Education, Culture, Sports, Science and Technology (MEXT), High Energy Accelerator Research Organization (KEK) and Japan Aerospace Exploration Agency (JAXA) in Japan, and the K. A. Wallenberg Foundation, the Swedish Research Council and the Swedish National Space Board in Sweden. Additional support for science analysis during the operations phase from the following agencies is also gratefully acknowledged: the Istituto Nazionale di Astrofisica in Italy and the Centre National d'Etudes Spatiales in France. This work was performed in part under DOE Contract DE-AC02-76SF00515. 

This research made use of Astropy,\footnote{\url{http://www.astropy.org}} a community-developed core Python package for Astronomy \citep{astropy:2013, astropy:2018}.

The authors would like to thank M.~Kerr for providing the updated pulsar ephemeris for PSR J1833$-$1034 and J. Li for providing the \textit{Fermi}-LAT spectrum of HESS J0632+057.
\end{acknowledgements}

%
%

\bibliographystyle{aa}
\bibliography{sample.bib}
\begin{appendix}
\onecolumn
\section{\textit{Swift} observations} \label{ap-xray}
\begin{table}[h]
\caption{\textit{Swift}-XRT observations of XMMU~J183245-0921539 in PC mode.}             
\label{table:xray}      
\centering      
\begin{minipage}{0.49\textwidth}    
\centering     
\begin{tabular}{c |c c}        
\hline\toprule      
OBS$\_$ID & MJD & Exposure (s)\\
\midrule
00010378001 & 58203 & 4707.4\\     
00010378002 & 58210 & 2414.9\\        
00010378003 & 58212 & 1920.4\\        
00010378004 & 58217 & 2230.1\\        
00010378005 & 58218 & 2622.2\\        
00010378006 & 58224 & 3860.9\\        
00010378007 & 58231 & 4849.7\\        
00010378008 & 58238 & 4585.0\\
00010378009 & 58245 & 4859.7\\        
00010378010 & 58252 & 4732.3\\        
00010378011 & 58259 & 4124.1\\        
00010378012 & 58266 & 4615.0\\        
00010378014 & 58280 & 4759.8\\     
00010378015 & 58287 & 4659.9\\
00034056001 & 57285 & 2599.7\\  
00034056002 & 57291 & 2826.9\\  
00034056003 & 57471 & 1795.6\\  
00034056004 & 57472 & 1168.7\\  
00034056005 & 57486 & 1158.8\\  
00034056006 & 57488 & 1478.7\\  
00034056007 & 57500 & 3281.0\\  
00034056008 & 57513 & 2197.6\\  
00034056009 & 57518 & 549.4\\  
00034056010 & 57529 & 159.8\\  
00034056011 & 57530 & 1858.0 \\  
00034056012 & 57538 & 641.8\\ 
00034056014 & 57541 & 1855.5\\  
00034056015 & 57555 & 1186.2\\  
00034056016 & 57557 & 626.8\\  
00034056017 & 57561 & 0.0 \\  
00034056018 & 57572 & 1445.9\\  
00034056019 & 57576 & 894.0\\  
00034056020 & 57578 & 317.2\\  
\bottomrule
\end{tabular}
\end{minipage} \hfill
\begin{minipage}{0.49\textwidth}
\centering   
\begin{tabular}{c |c c}    
\hline\toprule    
OBS$\_$ID & MJD & Exposure (s)\\
\midrule
00034056021 & 57584 & 3066.7\\ 
00034056022 & 57597 & 3596.1\\  
00034056023 & 57611 & 3728.4\\  
00034056024 & 57625 & 2122.7\\ 
00034056025 & 57626 & 357.1\\
00034056026 & 57631 & 472.0\\ 
00034056027 & 57639 & 1910.4\\ 
00034056028 & 57644 & 174.8 \\   
00034056029 & 57646 & 1423.5\\  
00034056030 & 57653 & 2250.1\\  
00034056031 & 57658 & 1311.1\\  
00034056032 & 57666 & 2554.7\\  
00034056034 & 57674 & 1256.1\\  
00034056035 & 57682 & 2130.2\\  
00034056036 & 57685 & 5.0\\  
00034056037 & 57688 & 2899.3\\  
00034056038 & 57694 & 3710.9\\  
00034056039 & 58165 & 1780.6\\  
00034056040 & 58179 & 1770.6\\  
00034056041 & 58193 & 1947.9\\  
00034056042 & 58207 & 2005.4\\  
\midrule
00034056031 & 57658 & 1311.1\\ 
00036174001 & 54524 & 2320.1\\  
00036174002 & 56974 & 252.2\\  
00036174004 & 57702 & 1028.9\\ 
00044299001 & 56242 & 524.4 \\ 
00044307001 & 56244 & 489.5 \\ 
00081776001 & 57468 & 1730.6\\  
00087484001 & 58016 & 4490.1\\  
00087484003 & 58177 & 474.5\\  
00087518004 & 58174 & 397.1\\  
\bottomrule

\end{tabular}
$\;$\\
$\;$\\
$\;$\\

\end{minipage}\hfill
\end{table}

\begin{table}[h]
\caption{UVOT upper limits at $95\%$ confidence level in erg cm$^{-2}$ s$^{-1}$ $\AA^{-1}$ }             
\label{table:uvot}      
\centering                          
\begin{tabular}{c |c c c c c}        
\hline\toprule      
Filter & V & U & UVM2 & UVW1 & UVW2   \\    
\midrule                       
   Flux  & $2.03 \cdot 10 ^{-17}$ & $1.76 \cdot 10 ^{-17}$ & $5.42 \cdot 10 ^{-17}$ & $2.80 \cdot 10 ^{-17}$ & $2.30 \cdot 10 ^{-17}$\\ 
\end{tabular}
\end{table}

\end{appendix}

\end{document}